\documentclass[11pt]{article}

\pdfoutput=1
\usepackage{verbatim}
\usepackage{amssymb}
\usepackage{xcolor}
\usepackage[centertags]{amsmath}
\usepackage[square, comma, sort&compress,numbers]{natbib}	
\usepackage{array,multirow}
\usepackage{graphicx}

\usepackage{bbold}
\usepackage{braket}
\usepackage{wrapfig}
\usepackage{float}
\usepackage{soul}
\usepackage{url}
\usepackage{multicol}
\numberwithin{equation}{section}
\usepackage{tikz}
\usepackage{tikz,pgf}
\usetikzlibrary{shapes}
\usetikzlibrary{calc}
\usetikzlibrary{decorations.pathmorphing}
\usetikzlibrary{decorations.pathreplacing,shapes.misc}
\usetikzlibrary{positioning}
\usetikzlibrary{arrows}
\usetikzlibrary{decorations.markings}
\usetikzlibrary{shadings}
\usetikzlibrary{intersections}
\usepackage{jheppub}
\usepackage{csquotes}
\numberwithin{equation}{section} \makeatletter
\@addtoreset{equation}{section}

\newcommand{\be}{\begin{equation}}
\newcommand{\ee}{\end{equation}}

\makeatletter
\def\@fpheader{\vspace{-.1cm}}
\makeatother

\title{On different approaches to integrable lattice models II
}

\author{Vladimir Belavin$^1$, Doron Gepner$^2$, J. Ramos Cabezas$^1$, Boris Runov$^{1}$}

\affiliation{$^1$Physics Department, Ariel University, Ariel 40700, Israel.}

\affiliation{$^2$Department of Particle Physics and Astrophysics, Weizmann Institute, Rehovot 76100, Israel.}

\emailAdd{vladimirbe@ariel.ac.il, doron.gepner@weizmann.ac.il, juanjose.ramoscab@msmail.ariel.ac.il, borisru@ariel.ac.il}

\abstract{This paper represents a continuation of our previous work, where the Bolzmann weights (BWs) for several Interaction-Round-the Face (IRF) lattice models were computed using their relation to rational conformal field theories. Here, we focus on deriving solutions for the Boltzmann weights of the Interaction-Round the Face lattice model, specifically the unrestricted face model, based on the $\mathfrak{su}(3)_k$ affine Lie algebra. The admissibility conditions are defined by the adjoint representation. We find the BWs by determining the quantum $R$ matrix of the $U_q(\mathfrak{sl} (3))$ quantum algebra in the adjoint representation and then applying the so-called Vertex-IRF correspondence. The Vertex-IRF correspondence defines the BWs of IRF models in terms of $R$ matrix elements.} 

\begin{document}
\maketitle
\flushbottom 

\section{Introduction}
This paper is a continuation of the previous work  \cite{Belavin:2023uqr}.
Here, we present the Boltzmann weights (BWs) of the unrestricted Interaction-Round the Face (IRF) lattice model based on the affine Lie algebra $\mathfrak{su}(3)_k$ and the $U_q(\mathfrak{sl}(3))$ quantum algebra, where the adjoint representation defines the admissibility conditions of the face configurations. The BWs are expressed in terms of the quantum $R$ matrix.

It is well-established that certain lattice models at criticality are closely connected to two-dimensional conformal field theories\footnote{Some examples include the Ising model \cite{Belavin:1984vu}, Yang-Lee edge singularity \cite{Cardy:1985yy}, the tricritical Ising model \cite{Blume:1971zza, Friedan:1984rv}, the three-state Potts model \cite{Potts:1951rk, Dotsenko:1984if}, the eight-vertex SOS model \cite{Andrews:1984af, Huse:1984mn}, and IRF models based on affine Lie algebras \cite{Date:1987zz,jimbo1987solvable,Jimbo:1987mu}.} (CFTs). In our previous paper \cite{Belavin:2023uqr}, we utilized this connection between CFT and restricted IRF models to determine the BWs for IRF models based on the affine Lie algebras $\mathfrak{su}(2)_k$ and $\mathfrak{su}(3)_k$, considering various levels $k$ and representations defining the admissibility conditions. We referred to that approach as the \enquote{CFT approach}. In this paper, we adopt a different approach, namely, the Vertex-IRF correspondence approach, to solve the aforementioned unrestricted IRF model. Ultimately, we aim to develop a general procedure to solve generic IRF models, which may allow us to explore some intriguing properties\footnote{For instance, those properties conjectured in \cite{Gepner:2020ajg, Belavin:2021uzv, Ramos:2022iun} regarding the fixed point theory of IRF models.} of these models and further connections with CFTs.

The model under study is defined on a two-dimensional square lattice, where the fluctuating variables residing on the lattice vertices (see Figure \ref{fig1}) belong to the set of integral weights of the algebra $\mathfrak{su}(3)_k$. The adjoint representation of the corresponding finite Lie algebra $\mathfrak{su}(3)$ will be used to define the admissibility conditions for face configurations. A comprehensive definition of the model will be provided in section \ref{description}. The BWs of the model satisfy the Yang-Baxter equation (YBE). In section \ref{virfc}, we discuss the Vertex-IRF correspondence and determine solutions for the BWs by finding the quantum $R$ matrix and applying the mentioned correspondence. All the non-zero $R$ matrix elements are listed in section \ref{jqr}. The derivation of the BWs for the model constitutes the main result of this paper. 
In section \ref{secRepTheory}, we explore the structure of our solution from the point of view of representation theory and compare our results to the naive CFT approach. Finally, we present our conclusions in section \ref{conclusions}, and in appendix \ref{generators}, we provide a description of the generators of $U_q(\mathfrak{sl}(3))$ in the adjoint representation.

\section{Description of the unrestricted IRF model} \label{description}
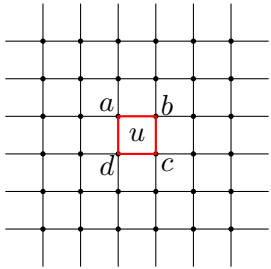
\begin{figure}[H]
    \centering

\begin{tikzpicture}[scale=0.5]
  \foreach \y in {1,2,3,4,5,6}
    \draw (0,\y) -- (7,\y);
  
  \foreach \x in {1,2,3,4,5,6}
    \draw (\x,0) -- (\x,7);

  \foreach \x in {1,2,3,4,5,6}
    \foreach \y in {1,2,3,4,5,6}
      \fill (\x,\y) circle (2pt);

  \draw[red, thick] (3,3) rectangle (4,4);

  \node at (2.7,4.3) {$a$};
  \node at (4.3,4.3) {$b$};
  \node at (4.3,2.7) {$c$};
  \node at (2.7,2.7) {$d$};
  \node at (3.5,3.5) {$u$};
\end{tikzpicture}
    \caption{Two-dimensional lattice.}
    \label{fig1}
\end{figure}

Here, we present the unrestricted IRF model based on $\mathfrak{su}(3)_k$. The model is defined on a two-dimensional square lattice, as illustrated in Figure \ref{fig1}, with fluctuating variables residing on the lattice vertices. A \textit{face} is formed by four nearest neighboring vertices (in Figure \ref{fig1}, a face with vertices $a, b, c,$ and $d$ is shown). The fluctuating variables take values from the set of \textit{integral weights} $P$ of $\mathfrak{su}(3)_k$

\begin{equation} \label{inw}
P=\left\{  a= a_0\Lambda_0+  a_1\Lambda_1+  a_2\Lambda_2= (a_0,a_1, a_2), \quad \quad a_i \in \mathbf{Z} \right\}.
\end{equation}
Let us introduce the weights of the adjoint representation of the corresponding finite algebra $\mathfrak{su}(3)$ as follows
\begin{equation} \label{wofadj}   
\begin{array}{llll}
     e_1= (1,1), & e_2= (-1,2), & e_3= (2,-1), & e_4= (0,0)_1, \\ e_5 = (0,0)_2, & e_6 = (1,-2), & e_7 = (-2,1), &e_8= (-1,-1).
\end{array}
\end{equation}
Their corresponding affine extensions (at level zero) are
\begin{equation}
\begin{array}{llll}
     \hat{e}_1= (-2, 1,1), & \hat{e}_2= (-1, -1,2), & \hat{e}_3= (-1, 2,-1), & \hat{e}_4= (0,0,0)_1, \\ \hat{e}_5 = (0,0,0)_2, & \hat{e}_6 = (1,1,-2,), & \hat{e}_7 = (1,-2,1), &\hat{e}_8= (2,-1,-1).
\end{array}
\end{equation}
For the unrestricted IRF model, we define the admissibility conditions as follows: A pair $(a,b)\in P$ is termed \textit{admissible} if
\begin{equation} \label{admiconun}
    b=a+\hat{e}_i, \quad \text{for some $i=1,2,...,8.$}
\end{equation}
 Now, let $(a,b,c,d)$ be the values of the North-West (NW), NE, SE, and SW corners of a face (e.g., the face shown in Figure \ref{fig1}). The face configuration $(a,b,c,d)$ is termed \textit{admissible} if the pairs $(a,b)$, $(a,d)$, $(b,c)$, and $(d,c)$ are all admissible. This definition of admissibility conditions is based on \cite{Jimbo:1987ra, Kuniba:1989br, Kuniba:1991yn}. For each face configuration, we assign a \textit{Boltzmann weight}
\begin{equation}
 \omega \left(\begin{matrix}  a & b \\ d & c 
\end{matrix}\bigg | u \right) .
\end{equation}
Thus, the BWs depend on the configuration $(a,b,c,d)$, as well as on the \textit{spetral parameter} $u$. For non-admissible face configurations, we set the Boltzmann weight to zero. For admissible face configurations, we require that the BWs satisfy the Yang-Baxter equation 
\begin{equation} \label{ibw}
\sum_g \omega \left( \begin{matrix}  a & b \\ f & g 
\end{matrix}\bigg | u +v\right)  \omega \left( \begin{matrix}  f & g \\ e & d 
\end{matrix}\bigg | u \right) \omega \left( \begin{matrix}  b & c \\ g & d 
\end{matrix}\bigg | v \right) = \sum_g \omega \left( \begin{matrix}  a & g \\ f & e
\end{matrix}\bigg | v\right)  \omega \left( \begin{matrix}  a & b \\ g & c 
\end{matrix}\bigg | u \right) \omega \left( \begin{matrix}  g & c \\ e & d 
\end{matrix}\bigg | u+v \right). 
\end{equation}
We have successfully found solutions to the YBE for the BWs in terms of the quantum $R$ matrix (which will be described in the next section). We have observed that for admissible face configurations ($a,b,c,d$) that do not involve the null weights $\hat{e}_4=(0,0,0)_1$ and $\hat{e}_5=(0,0,0)_2$, all the BWs are non-zero. However, for some admissible face configurations involving $\hat{e}_4$ or $\hat{e}_5$, certain BWs are zero. We interpret this result as an effect of the multiplicity of these two null vectors.

Before concluding this section, let us highlight an important difference between the definition (\ref{admiconun}) and the admissibility conditions of the restricted face model addressed in paper \cite{Belavin:2023uqr}. For the restricted IRF model, we defined an ordered pair ($a,b$) as admissible if $b$ appears in the tensor product of $a$ with the adjoint representation $(k-2, 1,1)$, that is, if $b \in a \otimes (k-2,1,1)$. It is evident that the condition (\ref{admiconun}) is less restrictive than the admissibility condition for restricted IRF models. Thus, the BWs of restricted IRF models can be obtained within the BWs of unrestricted models, and it is significantly important to note that, according to the authors of \cite{Jimbo:1987ra, Kuniba:1991yn}, the subset of \enquote{restricted} BWs satisfy the YBE among themselves.

In the forthcoming section, we will explain the Vertex-IRF correspondence through which we find solutions to the YBE for the BWs of the studied model.

\section{Vertex-IRF correspondence approach } \label{virfc}
The Vertex-IRF correspondence is a method that can be used to determine the BWs of unrestricted IRF models in terms of the quantum $\hat{R}$ matrix (for more details, see \cite{date1989one, jimbo1989introduction, Pasquier:1988nc}). In this section, we employ this method to find the BWs for the model described in the previous section. To begin, let us define the quantum $\hat{R}$ matrix. By its definition, the quantum $\hat{R}$ matrix satisfies the Yang-Baxter equation in the form
\begin{equation} \label{YBEeq2}
    ( I \otimes  \hat{R}(u)  ) (    \hat{R}(u+v) \otimes I       )  (  I\otimes \hat{R}(v)       ) =  (  \hat{R}(v)\otimes I )   (   I \otimes \hat{R}(u+v) )  ( \hat{R}(u) \otimes I ).
\end{equation}
Here, both sides of the equation act on the tensor product of three vector spaces: $V_1 \otimes V_2 \otimes V_3$. $I$ represents the identity operator, and $\hat{R}$ is given by
\begin{equation}
    \hat{R}= P \cdot  R,
\end{equation}
where $P$ is the transposition operator $P a \otimes b = b \otimes a$, and the operator $R$ (also called the quantum $R$ matrix) acts on the tensor product of two vector spaces. We denote its matrix elements as $R_{i,j}^{m,n}(u)$, namely,
\begin{equation} \label{matrixelR}
    R(u)(e_i \otimes e_j) = \sum_{m,n} R_{i,j}^{m,n}(u)\quad ( e_m \otimes e_n).
\end{equation}
For our purpose, $V_1, V_2, V_3$ will correspond to the vector space of the adjoint representation of $\mathfrak{su}(3)$, and hence $e_i$ (for $i= 1,2,...,8$) denotes the states (\ref{wofadj}).  It is common to write the YBE (\ref{YBEeq2}) in the form
\begin{equation} \label{YBEeqij}
R_{12}(u)   R_{13}(u+v)   R_{23}(v)=    R_{23}(v)   R_{13}(u+v)   R_{12}(u).
\end{equation}
In this form, $R_{ij}$ is an operator acting on the $i$th and $j$th components as $R$, and on the other component as the identity operator. We can directly write the YBE (\ref{YBEeq2}, \ref{YBEeqij}) in terms of $R$. For a given sextuplet ($i_1, i_2, i_3, f_1, f_2, f_3$), we have
\begin{equation} \label{YBEeq4}
  \sum_{s_1, s_2, s_3}   R_{s_2, s_3}^{f_1, f_2}(u) R_{i_1, s_1}^{s_2, f_3} (u+v)R_{i_2, i_3}^{s_3, s_1} (v) =  \sum_{s_1, s_2, s_3}R_{s_2, s_3}^{f_2, f_3}(v)R_{s_1, i_3}^{f_1, s_3}(u+v)R_{i_1, i_2}^{s_1, s_2} (u).
\end{equation}
As the adjoint representation contains 8 states, the matrix $R$ becomes a $64 \times 64$ matrix. However, many of its matrix elements are directly zero due to the property:
\begin{equation} \label{propertycgc}
    R_{i,j}^{k,l} = 0 \quad \text{if   $e_i+e_j\ne e_k+e_l$.}
\end{equation}
This property arises because the $R$ matrix can be expressed as a projector (see e.g.,  \cite{jimbo1989introduction, Kuniba:1989br}), the matrix elements of which are the Clebsch-Gordan coefficients (up to some factor), and it is known that the Clebsch-Gordan coefficients satisfy the property (\ref{propertycgc}).

The Vertex-IRF correspondence (see e.g., \cite{date1989one, jimbo1989introduction}) states that  the BWs of unrestricted IRF models can be determined in terms of the matrix elements\footnote{It is clear from the definition of $\hat{R}$ that $\hat{R}_{i,j}^{k,l}= R_{i,j}^{l,k}$.} of $\hat{R}$ as follows
\begin{equation} \label{vertexIRFeq}
   \omega  \left(\begin{matrix}  a & b \\ d & c 
\end{matrix}\bigg | u \right) =    \begin{cases} \hat{ R}_{i,j}^{k,l}(u) \quad \text{if $b-a=\hat{e}_k$, $c-b=\hat{e}_l$, $d-a= \hat{e}_i$, $c-d=\hat{e}_j$}  \\ 0 \quad \text{otherwise.} \end{cases}
\end{equation}
Indeed, if one defines the BWs according to this relation, one can see that equation (\ref{YBEeq4}) becomes the YBE (\ref{ibw}) for face models. In the following subsection, we will find the quantum $R$ matrix.

\subsection{Jimbo's method for the quantum $R$ matrix} \label{jqr}

In \cite{Jimbo:1985ua}, Jimbo provided a scheme for computing the quantum $R$ matrix, and in this subsection, we use this scheme. This scheme relies on certain elements. First, we need to introduce the $U_q(\mathfrak{sl} (3))$ quantum algebra, along with its generators in the adjoint representation. In this algebra, for each simple root $\alpha_a$ (for $a=1,2$), there are  three generators, namely, the Cartan, raising and lowering generators denoted respectively as follows
\begin{equation} \label{generatorshef}
    H_a, E_a, F_a.
\end{equation}
The quantum algebra $sl_q(3)$ is defined in terms of generators (\ref{generatorshef}) as follows
\begin{equation} \label{sl3qa}
    \begin{split}
       &  k_a= q^{H_a/2} ,\\ &
       k_a E_a=q E_a k_a, \quad k_a E_b= q^{-1/2}E_bk_a,\\&
       k_aF_a= q^{-1}F_a k_a,\quad k_a F_b= q^{1/2} F_bk_a, \\&
       [k_1, k_2]= [E_1,F_2]=[E_2,F_1]=0,\\&
       [E_a,F_a]=   \frac{k_a^2-k_a^{-2}}{q-q^{-1}} ,\\&
       E_a^2 E_b- (q+q^{-1})E_a E_bE_b+E_b E_a^2=0,\\&
       F_a^2 F_b-(q+q^{-1})F_aF_bF_a+F_bF_a^2=0,
    \end{split}
\end{equation}
where $a,b=1,2$. In \cite{Belavin:2023uqr}, we constructed the generators (\ref{generatorshef}) in the adjoint representation, satisfying the algebra (\ref{sl3qa}). In appendix \ref{generators}, we provide their explicit expressions. We will need the weights of the adjoint representation (\ref{wofadj}) in the Euclidean basis (these are eigenvectors of $H_1, H_2$ given by (\ref{apgenh1h2})). They are  given by
\begin{equation}
\begin{array}{ccc}
  e_1= ( 1,0,0,0,0,0,0,0  )^T,  &
 e_2= ( 0,1,0,0,0,0,0,0  )^T, &
 e_3= ( 0,0,0,0,0,1,0,0  )^T,\\
 e_4= ( 0,0,1,0,0,0,0,0  )^T,&
 e_5= ( 0,0,0,0,0,0,1,0  )^T,&
 e_6= ( 0,0,0,1,0,0,0,0  )^T,\\
 e_7= ( 0,0,0,0,0,0,0,1  )^T,&
 e_8= ( 0,0,0,0,1,0,0,0  )^T.
\end{array}
\end{equation}
Additionally, in this scheme, one will need the following two important elements (see equation  (2.2) of \cite{Jimbo:1985ua})
\begin{equation}
    k_0=  q^{(H_1+H_2)/2}, \quad  E_0= q^{  (H_1-H_2)/3 } \left(   F_2 F_1-q^{-1} F_1 F_2  \right).    
\end{equation}
Jimbo showed that the quantum $R$ matrix (\ref{matrixelR}, \ref{YBEeq4}) satisfies the following system of linear equations\footnote{We have noticed that work \cite{Ma:1990gg} partially overlaps with our discussion.} (for which, up to an overall factor, the solution is unique)
\begin{equation} \label{syseq1}
\begin{split}
   & R(u)   \left(  E_a \otimes  k_a^{-1}  +k_a \otimes E_a \right)= \left(    E_a \otimes k_a +k_a^{-1} \otimes E_a \right) R(u), \\&
      R(u)   \left(  F_a \otimes  k_a^{-1}  +k_a \otimes F_a \right)= \left(    F_a \otimes k_a +k_a^{
-1} \otimes F_a \right) R(u) ,\\ &
     [ R(u), H_a \otimes I+  I\otimes H_a]=0,\\&
      R(u) \left(   e^u  E_0 \otimes k_0 +k_0^{-1} \otimes E_0 \right) = \left(    e^{u}    E_0     \otimes k_0^{-1}   +k_0 \otimes E_0\right) R(u).
\end{split}
\end{equation} 
It is clear that, besides the spectral parameter $u$, the $R$ matrix also depends on the parameter $q$. To make identification with an IRF based on affine Lie algebra $\mathfrak{su}(n)_k$, we must set it to a specific root of unity:
\be
    q=e^{\frac{\pi i}{k+g}}\,,
\ee
where $g$ is the dual Coxeter number of the Lie algebra $\mathfrak{su}(n)$. 
Generally speaking, the representation theory of $U_q(\mathfrak{sl}(3))$ mirrors that of undeformed $\mathfrak{sl}(3)$ unless $q$ is a root of unity: for each finite-dimensional irreducible highest weight representation of $\mathfrak{sl}(3)$ there is an irreducible highest weight representation of $U_q(\mathfrak{sl}(3))$ of similar dimension. This is no longer true if $q$ is a root of unity. However, any particular finite-dimensional representation irreducible for generic $q$ exists and remains irreducible at roots of unity except for a finite number of values of $q$.
For example, the adjoint representation remains irreducible unless $q^6=1$.

By using the generators (\ref{generatorshef}) provided in appendix \ref{generators} and substituting them into (\ref{syseq1}), we have found solutions to this system of linear equations for the matrix elements $R_{i,j}^{k,l}(u)$ (for brevity, we denote $R_{i,j}^{k,l}(u) = R_{i,j}^{k,l}$). By using the relation (\ref{vertexIRFeq}), the matrix elements $R_{i,j}^{k,l}$ provide us with solutions for the BWs of the model studied in this paper. Here, we list the matrix elements we found, which constitute our main result. The solutions are parameterized by the factor $s_1 = R_{2,6}^{4,5}$, and we have verified that the following matrix elements satisfy YBE (\ref{YBEeq4}).
\clearpage

\begin{equation*}\label{matrixREl1}
\begin{split}  & R_{1,1}^{1,1}=R_{2,2}^{2,2}=R_{3,3}^{3,3}=R_{6,6}^{6,6}= R_{7,7}^{7,7}=R_{8,8}^{8,8}=\frac{s_1 e^{-u} \left(q^2 e^u-1\right)^2 \left(q^6 e^u-1\right)}{\left(q^2-1\right)^2 \left(q^2+1\right) \left(e^u-1\right)},\\&   
R_{1,2}^{1,2}=R_{1,3}^{1,3}=R_{2,1}^{2,1}=R_{2,7}^{2,7}=R_{3,1}^{3,1}=R_{3,6}^{3,6}=R_{6,3}^{6,3}=R_{6,8}^{6,8}=R_{7,2}^{7,2}=\\ & R_{7,8}^{7,8}=R_{8,6}^{8,6}=R_{8,7}^{8,7}=\frac{q s_1 e^{-u} \left(q^2 e^u-1\right) \left(q^6 e^u-1\right)}{\left(q^2-1\right)^2 \left(q^2+1\right)} , \\&
R_{1,2}^{2,1}=R_{1,3}^{3,1}=R_{1,4}^{4,1}=R_{2,7}^{7,2}=R_{3,6}^{6,3}=R_{4,8}^{8,4}=R_{6,8}^{8,6}= \\&R_{7,8}^{8,7}=\frac{s_1 \left(q^2 e^u-1\right) \left(q^6 e^u-1\right)}{\left(q^4-1\right) \left(e^u-1\right)}, \\&
R_{1,4}^{1,4}=R_{3,4}^{3,4}=R_{4,5}^{4,5}=R_{4,7}^{4,7}=R_{4,8}^{4,8}=R_{5,1}^{5,1}=R_{5,2}^{5,2}=R_{5,4}^{5,4}=R_{6,5}^{6,5} =\\& R_{8,5}^{8,5}=\frac{q^2 s_1 e^{-u} \left(e^u-1\right) \left(q^6 e^u-1\right)}{\left(q^2-1\right)^2 \left(q^2+1\right)}, \\&
R_{1,4}^{1,5}=R_{1,6}^{3,5}=R_{2,8}^{5,7}=R_{3,2}^{1,5}=R_{3,2}^{4,1}=R_{3,4}^{3,5}=R_{3,7}^{4,5}=R_{3,7}^{5,4}=R_{4,7}^{5,7}=\\& R_{4,8}^{5,8}=R_{5,1}^{4,1}=R_{5,2}^{4,2}=R_{6,2}^{4,5}=R_{6,2}^{5,4}=R_{6,5}^{6,4}=R_{6,7}^{5,8}=R_{6,7}^{8,4}=R_{7,1}^{4,2}=\\& R_{8,3}^{6,4}=R_{8,5}^{8,4}=0, \\&
R_{1,4}^{2,3}=R_{1,5}^{3,2}=R_{2,8}^{4,7}=R_{2,8}^{7,4}=R_{3,2}^{5,1}=R_{4,8}^{6,7}=R_{5,2}^{7,1}=R_{6,4}^{8,3}=\frac{\sqrt{q} s_1 \left(q^6 e^u-1\right)}{q^4-1},\\&
R_{1,4}^{3,2}=R_{1,6}^{3,4}=R_{1,6}^{4,3}=R_{6,5}^{8,3}=R_{6,7}^{8,5}=\frac{s_1 \left(q^6 e^u-1\right)}{\sqrt{q} \left(q^2-1\right)} ,\\&
R_{4,4}^{1,8}=\frac{q^2 s_1 e^{-u} \left(q^4-q^2 \left(e^u-1\right)-1\right)}{\left(q^2-1\right) \left(q^2+1\right)^2},\\&
R_{4,4}^{7,3}=\frac{s_1 \left(q^8 e^u-\left(q^2+1\right) q^2+1\right)}{q \left(q^2-1\right) \left(q^2+1\right)^2},\\&
R_{1,5}^{1,4}=R_{3,5}^{3,4}=R_{3,5}^{4,3}=R_{4,1}^{5,1}=R_{4,2}^{2,5}=R_{4,2}^{5,2}=R_{4,4}^{4,5}=R_{4,4}^{5,4}=R_{5,5}^{4,5}= \\& R_{5,5}^{5,4}=R_{5,7}^{4,7}=R_{5,7}^{7,4}=R_{5,8}^{4,8}=R_{6,4}^{5,6}=R_{6,4}^{6,5}=R_{8,4}^{8,5}=-\frac{q^3 s_1}{q^2+1},\\&
R_{1,5}^{1,5}=R_{3,5}^{3,5}=R_{4,1}^{4,1}=R_{4,2}^{4,2}=R_{5,7}^{5,7}=R_{5,8}^{5,8}=R_{6,4}^{6,4}=\\& R_{8,4}^{8,4}=\frac{q^2 s_1 e^{-u} \left(q^2 e^u-1\right) \left(q^4 e^u-1\right)}{\left(q^2-1\right)^2 \left(q^2+1\right)},\\&
R_{1,5}^{2,3}=R_{7,4}^{8,2}=\frac{q^{3/2} s_1 \left(q^4 \left(\left(q^2+1\right) e^u-1\right)-1\right)}{q^4-1},\\&
R_{1,5}^{5,1}=R_{5,8}^{8,5}=\frac{s_1 \left(q^2+e^u \left(q^8 e^u-\left(q^4+2\right) q^2+1\right)\right)}{\left(q^4-1\right) \left(e^u-1\right)},
\end{split}
\end{equation*}

\begin{equation*} \label{matrixREl2}
\begin{split}
 & R_{1,6}^{1,6}=R_{2,8}^{2,8}=R_{3,2}^{3,2}=R_{6,7}^{6,7}=R_{7,1}^{7,1}=R_{8,3}^{8,3}=\frac{q s_1 e^{-u} \left(e^u-q^2\right) \left(q^6 e^u-1\right)}{\left(q^2-1\right)^2 \left(q^2+1\right)},\\&
  R_{1,6}^{6,1}=R_{1,7}^{7,1}=R_{2,8}^{8,2}=R_{3,8}^{8,3}=\frac{s_1 e^u \left(q^6 e^u-1\right)}{\left(q^2+1\right) \left(e^u-1\right)},\\&
  R_{1,7}^{1,7}=R_{2,3}^{2,3}=R_{3,8}^{3,8}=R_{6,1}^{6,1}=R_{7,6}^{7,6}=R_{8,2}^{8,2}=\frac{q^3 s_1 e^{-u} \left(e^u-1\right) \left(q^4 e^u-1\right)}{\left(q^2-1\right)^2 \left(q^2+1\right)},\\& 
R_{1,8}^{1,8}=R_{2,6}^{2,6}=R_{3,7}^{3,7}=R_{6,2}^{6,2}=R_{7,3}^{7,3}=R_{8,1}^{8,1}=\frac{q^2 s_1 e^{-u} \left(e^u-q^2\right) \left(q^4 e^u-1\right)}{\left(q^2-1\right)^2 \left(q^2+1\right)},\\&
  R_{1,8}^{6,2}=R_{1,8}^{7,3}=R_{2,6}^{8,1}=R_{3,7}^{8,1}=-\frac{q^3 s_1 e^u}{q^2+1},\\&
 R_{1,8}^{8,1}=\frac{s_1 e^u \left(q^2 \left(\left(q^4+1\right) e^u-1\right)-1\right)}{\left(q^2+1\right) \left(e^u-1\right)},\\&
R_{2,1}^{1,2}=R_{3,1}^{1,3}=R_{5,1}^{1,5}=R_{6,3}^{3,6}=R_{7,2}^{2,7}=R_{8,5}^{5,8}=R_{8,6}^{6,8}=R_{8,7}^{7,8}=\frac{s_1 e^{-u} \left(q^2 e^u-1\right) \left(q^6 e^u-1\right)}{\left(q^4-1\right) \left(e^u-1\right)},\\&
R_{2,3}^{1,4}=\frac{q^{9/2} s_1}{q^2+1}, \quad R_{2,3}^{1,5}=\frac{q^{3/2} s_1 e^{-u} \left(q^2 e^u-1\right)}{q^4-1},\quad R_{2,3}^{4,1}=\frac{q^{9/2} s_1 \left(q^2 e^u-1\right)}{q^4-1}, \\&
R_{2,3}^{3,2}=R_{3,2}^{2,3}=R_{3,4}^{4,3}=R_{4,5}^{5,4}=R_{4,7}^{7,4}=R_{5,2}^{2,5}=R_{5,4}^{4,5}=R_{6,5}^{5,6}=R_{6,7}^{7,6}=\\& R_{7,6}^{6,7}=\frac{s_1 \left(q^6 e^u-1\right)}{\left(q^2+1\right) \left(e^u-1\right)},\\&
R_{2,4}^{2,4}=R_{4,3}^{4,3}=R_{4,6}^{4,6}=R_{7,4}^{7,4}=\frac{q^2 s_1 e^{-u} \left(e^u \left(q^6 e^u-2 q^4+q^2-1\right)+1\right)}{\left(q^2-1\right)^2 \left(q^2+1\right)},\\&
R_{2,4}^{2,5}=R_{2,4}^{5,2}=R_{4,3}^{5,3}=R_{4,5}^{5,5}=R_{4,6}^{5,6}=R_{4,6}^{6,5}=R_{5,4}^{5,5}=R_{7,4}^{7,5}=\frac{q s_1}{q^2+1},\\&
R_{2,4}^{4,2}=R_{4,6}^{6,4}=\frac{s_1 \left(q^2 e^u-1\right) \left(q^4 \left(\left(q^2+1\right) e^u-1\right)-1\right)}{\left(q^4-1\right) \left(e^u-1\right)},\\&
R_{2,5}^{1,7}=R_{8,4}^{7,6}=\frac{q^{3/2} s_1 e^{-u} \left(q^2 \left(\left(q^4+1\right) e^u-1\right)-1\right)}{\left(q^2-1\right) \left(q^2+1\right)^2},\\&
R_{2,5}^{2,4}=R_{4,5}^{4,4}=R_{5,3}^{3,4}=R_{5,3}^{4,3}=R_{5,4}^{4,4}=R_{5,6}^{4,6}=R_{7,5}^{4,7}=R_{7,5}^{7,4}=\frac{q^5 s_1}{q^2+1}, \\&
R_{2,5}^{2,5}=R_{5,3}^{5,3}=R_{5,6}^{5,6}=R_{7,5}^{7,5}=\frac{q^2 s_1 e^{-u} \left(q^2 e^u \left(q^4 \left(e^u-1\right)+q^2-2\right)+1\right)}{\left(q^2-1\right)^2 \left(q^2+1\right)},\\& 
 R_{3,2}^{1,4}=R_{3,4}^{1,6}=R_{4,1}^{3,2}=R_{5,1}^{2,3}=R_{5,7}^{2,8}=R_{8,3}^{5,6}=R_{8,3}^{6,5}=R_{8,5}^{6,7}=\frac{q^{3/2} s_1 e^{-u} \left(q^6 e^u-1\right)}{q^4-1},\\&
 R_{3,5}^{1,6}=R_{8,4}^{6,7}=R_{8,5}^{7,6}=\frac{q^{5/2} s_1 e^{-u} \left(q^6 e^u-1\right)}{\left(q^2-1\right) \left(q^2+1\right)^2},\\&
  R_{3,5}^{5,3}=R_{5,7}^{7,5}=\frac{s_1 \left(q^2 \left(e^u \left(q^4 \left(\left(q^2+1\right) e^u-2\right)-2\right)+1\right)+1\right)}{\left(q^4-1\right) \left(e^u-1\right)},\\&
R_{4,1}^{1,4}=R_{8,4}^{4,8}=\frac{s_1 e^{-u} \left(q^2 e^u \left(q^6+q^4 \left(e^u-2\right)-1\right)+1\right)}{\left(q^4-1\right) \left(e^u-1\right)},\\&
\end{split}    
\end{equation*}

\begin{equation*} \label{matrixREl3}
\begin{split}
&R_{4,1}^{2,3}=R_{5,6}^{3,8}=\frac{\sqrt{q} s_1 e^{-u} \left(q^2 \left(\left(q^4+1\right) e^u-1\right)-1\right)}{q^4-1},\\&
R_{4,2}^{2,4}=R_{6,4}^{4,6}=\frac{s_1 e^{-u} \left(q^2 \left(e^u \left(q^4 \left(\left(q^2+1\right) e^u-2\right)-2\right)+1\right)+1\right)}{\left(q^4-1\right) \left(e^u-1\right)},\\&
R_{4,4}^{2,6}=R_{5,5}^{3,7}=\frac{q s_1 e^{-u} \left(q^2 \left(q^6+2 q^4+1\right) e^u-\left(q^2+1\right)^2\right)}{\left(q^2-1\right) \left(q^2+1\right)^2},\\&
R_{4,4}^{4,4}=\frac{s_1 e^{-u} \left(-q^2+\left(3 q^2-1\right) \left(q^4+1\right) e^u+q^6 e^{2 u} \left(q^4+2 q^2 \sinh (u)-2 q^2-2\right)\right)}{\left(q^2-1\right)^2 \left(q^2+1\right) \left(e^u-1\right)} ,\\&
 R_{4,4}^{6,2}=R_{5,5}^{7,3}=\frac{q s_1 \left(\left(q^2+1\right)^2 q^4 e^u-\left(q^4+2\right) q^2-1\right)}{\left(q^2-1\right) \left(q^2+1\right)^2},\\&
R_{4,5}^{3,7}=R_{5,4}^{3,7}=\frac{q^2 s_1 e^{-u} \left(\left(q^4-q^2+1\right) e^u-1\right)}{q^4-1},\\&
R_{4,7}^{2,8}=R_{5,1}^{3,2}=R_{6,7}^{4,8}=R_{7,1}^{2,5}=R_{7,1}^{5,2}=\frac{\sqrt{q} s_1 e^{-u} \left(q^6 e^u-1\right)}{q^2-1},\\&
R_{4,2}^{7,1}=R_{4,8}^{7,6}=R_{5,8}^{6,7}=\frac{q^{3/2} s_1 \left(q^6 e^u-1\right)}{\left(q^2-1\right) \left(q^2+1\right)^2},\\&
R_{4,3}^{3,4}=R_{7,4}^{4,7}=\frac{s_1 \left(q^4 \left(\left(q^2+1\right) e^u-1\right)-1\right)}{\left(q^2+1\right) \left(e^u-1\right)},\\&
R_{4,3}^{6,1}=R_{5,8}^{7,6}=\frac{q^{5/2} s_1 \left(q^4 \left(\left(q^2+1\right) e^u-1\right)-1\right)}{\left(q^2-1\right) \left(q^2+1\right)^2},\\&
R_{4,4}^{3,7}=-\frac{q s_1 e^{-u} \left(q^2+\left(q^4-q^2-1\right) e^u\right)}{\left(q^2-1\right) \left(q^2+1\right)^2},\\&
R_{5,3}^{3,5}=R_{7,5}^{5,7}=\frac{s_1 e^{-u} \left(q^2 e^u-1\right) \left(q^2 \left(\left(q^4+1\right) e^u-1\right)-1\right)}{\left(q^4-1\right) \left(e^u-1\right)},\\&
R_{5,5}^{5,5}=\frac{s_1 e^{-u} \left(e^u \left(q^2 \left(2 \left(q^2+1\right)+e^u \left(q^8+q^6 \left(e^u-3\right)+q^4-3 q^2+1\right)\right)-1\right)-q^2\right)}{\left(q^2-1\right)^2 \left(q^2+1\right) \left(e^u-1\right)},\\&
R_{5,5}^{1,8}=-\frac{q^2 s_1 e^{-u} \left(q^8 e^u-\left(q^2+1\right) q^2+1\right)}{\left(q^2-1\right) \left(q^2+1\right)^2},\\&
R_{5,5}^{2,6}=\frac{q^3 s_1 e^{-u} \left(\left(-q^8+q^6+q^4\right) e^u-1\right)}{\left(q^2-1\right) \left(q^2+1\right)^2},\\&
R_{6,1}^{1,6}=R_{7,1}^{1,7}=R_{8,2}^{2,8}=R_{8,3}^{3,8}=\frac{s_1 e^{-u} \left(q^6 e^u-1\right)}{\left(q^2+1\right) \left(e^u-1\right)},\\&
R_{6,2}^{2,6}=R_{7,3}^{3,7}=s_1 \left(\frac{q^6-1}{\left(q^2+1\right) \left(e^u-1\right)}+e^{-u}\right),
\end{split}
\end{equation*}

\begin{alignat*}{100}
\label{matrixREl4}
& R_{1,4}^{5,1}=R_{4,8}^{8,5}=\frac{s_1 \left(1-q^6 e^u\right)}{q \left(q^4-1\right)} ,  && R_{1,5}^{4,1}=R_{5,8}^{8,4}=-\frac{q^3 s_1 \left(q^2 e^u-1\right)}{q^4-1},\\
& R_{1,6}^{5,3}=\frac{s_1 \left(1-q^6 e^u\right)}{q^{3/2} \left(q^2-1\right)}, &&
   R_{1,7}^{2,4}=-q^{5/2} s_1, \\
&  R_{1,7}^{2,5}=R_{1,7}^{5,2}=\frac{q^{3/2} s_1 \left(q^4 e^u-1\right)}{q^2-1}, &&
R_{1,7}^{4,2}=-\frac{q^{5/2} s_1 \left(q^2 e^u-1\right)}{q^2-1},\\
& R_{1,8}^{2,6}=R_{7,3}^{8,1}=\frac{q^5 s_1 \left(e^u-q^2\right)}{q^4-1}, && 
  R_{1,8}^{3,7}=R_{6,2}^{8,1}=\frac{q s_1 \left(q^4 e^u-1\right)}{q^4-1} ,    \\
  & R_{1,8}^{4,4}=-\frac{q^2 s_1 \left(q^2 e^u-1\right)}{q^2-1},   &&
  R_{1,8}^{4,5}=R_{1,8}^{5,4}=R_{3,7}^{5,5}=\frac{q s_1 \left(q^4 e^u-1\right)}{q^2-1},\\
& R_{1,8}^{5,5}=\frac{s_1 \left(1-q^4 e^u\right)}{q^2-1},&&  R_{2,3}^{5,1}=\frac{q^{3/2} s_1}{q^2+1},    \\
& R_{2,4}^{1,7}=\frac{q^{5/2} s_1 e^{-u} \left(e^u-1\right)}{\left(q^2-1\right) \left(q^2+1\right)^2}, && R_{2,4}^{7,1}=\frac{s_1 \left(1-q^6 e^u\right)}{\sqrt{q} \left(q^2-1\right) \left(q^2+1\right)^2},\\
 & R_{2,5}^{4,2}=R_{5,6}^{6,4}=\frac{q^5 s_1 \left(q^2 e^u-1\right)}{q^4-1}, && R_{2,5}^{7,1}=R_{3,4}^{6,1}=\frac{q^{5/2} s_1 \left(q^6 e^u-1\right)}{\left(q^2-1\right) \left(q^2+1\right)^2},\\
 & R_{2,5}^{5,2}=R_{5,6}^{6,5}=\frac{s_1 \left(q^2 \left(\left(q^4+1\right) e^u-1\right)-1\right)}{\left(q^2+1\right) \left(e^u-1\right)}, && R_{2,6}^{1,8}=R_{8,1}^{7,3}=\frac{q s_1 e^{-u} \left(e^u-q^2\right)}{q^4-1},\\
  & R_{2,6}^{6,2}=R_{3,7}^{7,3}=\frac{s_1 e^u \left(q^4 \left(\left(q^2+1\right) e^u-1\right)-1\right)}{\left(q^2+1\right) \left(e^u-1\right)}, && R_{2,6}^{4,5}= R_{2,6}^{5,4}=R_{4,4}^{5,5}=s_1, \\
& R_{2,6}^{3,7}=R_{6,2}^{7,3}=\frac{s_1}{q^2+1}, && 
R_{2,6}^{4,4}=\frac{q^3 s_1 \left(q^2 e^u-1\right)}{q^2-1}, \\
& R_{2,6}^{5,5}=-\frac{s_1}{q}, \quad R_{6,1}^{3,5}=-\frac{q^{3/2} s_1 e^{-u} \left(q^2 e^u-1\right)}{q^2-1}, && 
R_{2,6}^{7,3}=\frac{s_1 \left(1-q^4 e^u\right)}{q^4-1} ,\\
& R_{2,8}^{7,5}=\frac{s_1 \left(1-q^6 e^u\right)}{\sqrt{q} \left(q^4-1\right)}, && 
R_{3,4}^{5,3}=R_{4,7}^{7,5}=\frac{q s_1 \left(q^6 e^u-1\right)}{q^4-1} , \\
& R_{3,5}^{6,1}=-\frac{q^{11/2} s_1 \left(e^u-1\right)}{\left(q^2-1\right) \left(q^2+1\right)^2} , && 
R_{3,7}^{1,8}=R_{8,1}^{6,2}=\frac{q^3 s_1 e^{-u} \left(q^4 e^u-1\right)}{q^4-1}, \\
& R_{3,7}^{2,6}=R_{7,3}^{6,2}=\frac{q^6 s_1}{q^2+1} , && 
R_{3,7}^{4,4}=-q^3 s_1, \\
& R_{3,7}^{6,2}=\frac{q^4 s_1 \left(q^2-e^u\right)}{q^4-1}, && 
R_{3,8}^{4,6}=-\frac{q^{7/2} s_1}{q^2+1}, \\
& R_{3,8}^{5,6}=R_{3,8}^{6,5}=\frac{q^{5/2} s_1 \left(q^4 e^u-1\right)}{q^4-1}, && 
R_{3,8}^{6,4}=-\frac{q^{7/2} s_1 \left(q^2 e^u-1\right)}{q^4-1},\\
& R_{4,1}^{1,5}=R_{8,4}^{5,8}=-\frac{q^3 s_1 e^{-u} \left(q^2 e^u-1\right)}{q^4-1}, && 
R_{4,2}^{1,7}=-\frac{q^{9/2} s_1 e^{-u} \left(e^u-1\right)}{\left(q^2-1\right) \left(q^2+1\right)^2}, \\
& R_{4,3}^{1,6}=R_{5,2}^{1,7}=\frac{q^{3/2} s_1 e^{-u} \left(q^6 e^u-1\right)}{\left(q^2-1\right) \left(q^2+1\right)^2}, && 
R_{4,3}^{3,5}=R_{7,4}^{5,7}=\frac{q s_1 e^{-u} \left(q^2 e^u-1\right)}{q^4-1},
\end{alignat*}

\begin{equation}
\begin{aligned}
\label{matrixREl5}
& R_{4,4}^{8,1}=\frac{s_1 \left(\left(q^4-q^2-1\right) q^4 e^u+1\right)}{\left(q^2-1\right) \left(q^2+1\right)^2}, && 
R_{4,5}^{1,8}=R_{5,4}^{1,8}=\frac{q s_1 e^{-u} \left(q^6 e^u+q^4-q^2-1\right)}{\left(q^2-1\right) \left(q^2+1\right)^2},\\
& R_{4,5}^{2,6}=R_{5,4}^{2,6}=\frac{q^2 s_1 e^{-u} \left(2 q^6 e^u-q^2-1\right)}{\left(q^2-1\right) \left(q^2+1\right)^2},&& 
R_{4,5}^{6,2}=R_{5,4}^{6,2}=\frac{q^2 s_1 \left(q^4 \left(e^u-1\right)+q^2-1\right)}{q^4-1},\\
& R_{4,5}^{7,3}=R_{5,4}^{7,3}=\frac{q^2 s_1 \left(\left(q^6+q^4\right) e^u-2\right)}{\left(q^2-1\right) \left(q^2+1\right)^2},&& 
R_{4,5}^{8,1}=R_{5,4}^{8,1}=\frac{q^3 s_1 \left(q^2 \left(q^4+q^2-1\right) e^u-1\right)}{\left(q^2-1\right) \left(q^2+1\right)^2},\\
& R_{4,6}^{3,8}=\frac{q^{3/2} s_1 e^{-u} \left(e^u-1\right)}{q^4-1},&& 
R_{4,6}^{8,3}=\frac{s_1 \left(1-q^6 e^u\right)}{q^{3/2} \left(q^4-1\right)},\\
& R_{4,7}^{8,2}=R_{5,6}^{8,3}=\frac{q^{3/2} s_1 \left(q^6 e^u-1\right)}{q^4-1},&& 
R_{5,1}^{1,4}=R_{8,5}^{4,8}=-\frac{q^3 s_1 e^{-u} \left(q^6 e^u-1\right)}{q^4-1},\\
& R_{5,2}^{2,4}=R_{6,5}^{4,6}=\frac{q s_1 e^{-u} \left(q^6 e^u-1\right)}{q^4-1},&& 
R_{5,3}^{1,6}=-\frac{q^{9/2} s_1 e^{-u} \left(q^6 e^u-1\right)}{\left(q^2-1\right) \left(q^2+1\right)^2},\\
& R_{5,3}^{6,1}=\frac{q^{15/2} s_1 \left(e^u-1\right)}{\left(q^2-1\right) \left(q^2+1\right)^2} ,&& 
R_{5,5}^{4,4}=R_{7,3}^{4,5}=R_{7,3}^{5,4}=q^4 s_1 ,\\
& R_{5,5}^{6,2}=-\frac{q^5 s_1 \left(q^4-q^2 \left(e^u-1\right)-1\right)}{\left(q^2-1\right) \left(q^2+1\right)^2},&& 
R_{5,5}^{8,1}=\frac{q^4 s_1 \left(q^2+\left(q^4-q^2-1\right) e^u\right)}{\left(q^2-1\right) \left(q^2+1\right)^2},\\
& R_{5,7}^{8,2}=-\frac{q^{9/2} s_1 \left(e^u-1\right)}{q^4-1},&& 
R_{6,1}^{3,4}=R_{6,1}^{4,3}=\frac{\sqrt{q} s_1 e^{-u} \left(q^4 e^u-1\right)}{q^2-1},\\
& R_{6,2}^{1,8}=R_{7,3}^{1,8}=R_{8,1}^{2,6}=R_{8,1}^{3,7}=-\frac{q^3 s_1 e^{-u}}{q^2+1},&& 
R_{6,2}^{3,7}=\frac{q^2 s_1 e^{-u} \left(q^2-e^u\right)}{q^4-1},\\
& R_{6,2}^{4,4}=R_{8,1}^{4,5}=R_{8,1}^{5,4}=\frac{q s_1 e^{-u} \left(q^4 e^u-1\right)}{q^2-1},&& 
R_{6,1}^{5,3}=-q^{3/2} s_1, \quad R_{6,2}^{5,5}=-q s_1, \\
& R_{6,4}^{3,8}=-\frac{q^{7/2} s_1 e^{-u} \left(e^u-1\right)}{q^4-1}, && 
R_{6,5}^{3,8}=R_{7,4}^{2,8}=\frac{\sqrt{q} s_1 e^{-u} \left(q^6 e^u-1\right)}{q^4-1},\\
& R_{7,1}^{2,4}=-\frac{q^{3/2} s_1 e^{-u} \left(q^6 e^u-1\right)}{q^2-1},&& 
R_{7,3}^{2,6}=-\frac{q^4 s_1 e^{-u} \left(q^4 e^u-1\right)}{q^4-1},\\
& R_{7,3}^{4,4}=-q^5 s_1 ,\quad R_{7,6}^{8,4}=\frac{q^{7/2} s_1 \left(q^2 e^u-1\right)}{q^2-1}, &&   R_{7,3}^{5,5}=\frac{q s_1 e^{-u} \left(q^2 e^u-1\right)}{q^2-1} , \\
& R_{7,5}^{2,8}=-\frac{q^{7/2} s_1 e^{-u} \left(q^6 e^u-1\right)}{q^4-1} ,&& 
 R_{7,5}^{8,2}=\frac{q^{13/2} s_1 \left(e^u-1\right)}{q^4-1}, \\
& R_{7,6}^{4,8}=q^{7/2} s_1, \quad R_{7,6}^{8,5}=\sqrt{q} s_1 , && 
R_{7,6}^{5,8}=\frac{\sqrt{q} s_1 e^{-u} \left(q^2 e^u-1\right)}{q^2-1}, \\
& R_{8,1}^{1,8}=\frac{s_1 e^{-u} \left(q^4 \left(\left(q^2+1\right) e^u-1\right)-1\right)}{\left(q^2+1\right) \left(e^u-1\right)}, && 
R_{8,1}^{4,4}=-\frac{q^2 s_1 e^{-u} \left(q^4 e^u-1\right)}{q^2-1}, \\
& R_{8,1}^{5,5}=-\frac{q^2 s_1 e^{-u} \left(q^2 e^u-1\right)}{q^2-1}, && 
R_{8,2}^{4,7}=R_{8,2}^{7,4}=\frac{q^{3/2} s_1 e^{-u} \left(q^4 e^u-1\right)}{q^4-1} ,\\
& R_{8,2}^{5,7}=-\frac{q^{5/2} s_1 e^{-u} \left(q^2 e^u-1\right)}{q^4-1} ,&&  
R_{8,2}^{7,5}=-\frac{q^{5/2} s_1}{q^2+1},   R_{8,3}^{4,6}=-\frac{q^{5/2} s_1 e^{-u} \left(q^6 e^u-1\right)}{q^4-1}. \\
\end{aligned}
\end{equation}

\section{Structure of the solution and connection to RCFTs}
\label{secRepTheory}
Earlier, we discussed the approach to construction of integrable IRF models based on the fusion rules of a rational conformal field theory. As mentioned in the previous paper, this approach cannot be used for $U_q(\mathfrak{sl}(3))$ in the adjoint representation for generic $q$ due to the presence of the multiplicities.

Let $\rho^{(q)}_{\lambda}$ be an irreducible highest weight representation of the quantum group $U_q(\mathfrak{sl}(n))$,
$V_{\lambda}$ corresponding irreducible module, and
$\hat{\rho}^{(q)}_{\lambda,u}$ corresponding evaluation representation of the affine quantum group $U_q(\hat{\mathfrak{sl}}(n))$.
Let
\begin{equation}
\label{uniR}
    \mathcal{R} \in U_q(\hat{\mathfrak{g}}) \times U_q(\hat{\mathfrak{g}})
\end{equation}
be the universal $R$ matrix of the affine quantum group
$U_q(\hat{\mathfrak{sl}}(n))$, and $R_{\lambda\lambda}$ its representation:
\be
    R_{\lambda\lambda}(u)=\hat{\rho}^{(q)}_{\lambda,u}\otimes\hat{\rho}^{(q)}_{\lambda,0}(\mathcal{R}).
\ee
The matrix $R_{\lambda\lambda}$ is a solution to Yang-Baxter equation (\ref{YBEeq4}), and
\be
    \hat{R}_{\lambda\lambda}=P\cdot R_{\lambda\lambda}
\ee
solves (\ref{YBEeq2}) respectively.
The tensor product of representation spaces $V_{\lambda}\times V_{\lambda}$ can be decomposed into a direct sum of irreducible $U_q(\mathfrak{sl}(n))$ modules:
\be
\label{tensorProd}
    V_{\lambda}\times V_{\lambda}=\bigoplus\limits_{j=1}^{N} V_{\lambda_j}.
\ee
Since $\mathcal{R}$ is the universal $R$-matrix, the matrix $\hat{R}(u)$ must satisfy 
\be
    [\hat{R}(u),\rho^{(q)}_{\lambda}\otimes\rho^{(q)}_{\lambda}(\Delta(X))]=0\,,\;\forall X \in U_q(\mathfrak{sl}(n))\,,
\ee
where $\Delta(X)$ denotes the coproduct of $X$ in $U_q(\mathfrak{sl}(n))$.
Thus, the matrix $\hat{R}$  commutes with all the generators of the ``horizontal'' quantum subgroup $U_q(\mathfrak{sl}(n))$, in particular with the generators of the Cartan subalgebra, which implies it preserves the weight of any vector it acts upon. Consequently, any highest weight vector is mapped to another highest weight vector of the same weight. Therefore, at any value of the spectral parameter, the matrix $\hat{R}(u)$ can be decomposed into a sum of projectors onto irreducible submodules of $U_q(\mathfrak{sl}(n))$:
\be
\label{hatRsum}
    \hat{R}(u)=\sum_{j=1}^{N}f_{\lambda_j}(u) P^{(\lambda_j)}(u).
\ee
There could be two kinds of multiplicities in the decomposition (\ref{hatRsum}). Firstly, an irreducible representation can appear multiple times in the decomposition (\ref{tensorProd}) of the tensor product of the representation spaces. Secondly, some of the eigenvalues $f_{\lambda_k}(u)$ might coincide. Both kinds of multiplicities are present in the case of $U_q(\hat{sl}(3))$ at generic level $k$ considered in this work.

In the ``CFT'' approach, we recover $\hat{R}(u)$ from its UV limit. If $\lambda_j$ occurs only once in (\ref{tensorProd}), then $P^{(\lambda_j)}$ doesn't depend on $u$. We can, therefore, find the corresponding projector in the UV limit as
\be
\label{projectorUV}
    P^{(\lambda_j)}=\prod_{i\neq j}\frac{\hat{R}(\infty)-f_{\lambda_i}(\infty)}{f_{\lambda_j}(\infty)-f_{\lambda_i}(\infty)}
\ee
and substitute it into (\ref{hatRsum}) to get a manageable ansatz for an R matrix. On the other hand, if there are multiple submodules of the same highest weight in (\ref{tensorProd}), then the decomposition (\ref{tensorProd}) is not unique as there is an arbitrary choice of basis on the subspace of this weight. Consequently, the $R$ matrix might mix different copies of $V_{\lambda_j}$. So, we can write the R matrix as
\be
    \hat{R}=\sum_{m=1}^{N}\sum_{i,j=1}^{\#\lambda_m}
    M^{(\lambda_m)}_{ij}(u)P_{ij}^{(\lambda_m)},
\ee
where $P_{ii}^{(\lambda_m)}$ is a projector onto $i$-th copy of $V_{\lambda_m}$ in some fixed basis independent of the spectral parameter, $P_{ij}^{(\lambda_m)}$ is an operator which maps $V_{\lambda_m}^{(i)}$ to $V_{\lambda_m}^{(j)}$ and commutes with all the generators, $\#\lambda_m$ is the multiplicity of weight $\lambda_m$ in the decomposition (\ref{tensorProd}), and functions $M_{ij}^{(\lambda_m)}(u)$ capture the dependence on the spectral parameter.

The second kind of multiplicity is easier to deal with. Even though projectors onto the corresponding irreducible submodules cannot be obtained directly from the UV limit of the R-matrix using (\ref{projectorUV}), they are still defined unambiguously and do not depend on $u$. Moreover, (\ref{projectorUV}) can be generalised as
\be
    P_{f}=\sum_{f_{\lambda_i}(\infty)=f}P^{(\lambda_i)}=\prod_{f_{\lambda_i}(\infty) \neq f}\frac{\hat{R}(\infty)-f_{\lambda_i}(\infty)}{f-f_{\lambda_i}(\infty)}
\ee
to give the projector $P_{f}$ onto the eigenspace of the $R$ matrix with eigenvalue $f$, which is a direct sum of irreducible modules with that eigenvalue.
Based on the example of $U_q(sl(3))$ in the adjoint representation, we can speculate that such coincidences happen due to the symmetries of the weight system, and therefore, the eigenvalues coinciding in the UV limit stay equal for all values of the spectral parameter, in which case the knowledge of the projectors $P_{f}$ would be sufficient for the purpose of constructing the $R$ matrix.

Indeed,  we can rewrite the last equation of (\ref{syseq1}) as follows
\be
\label{affineRootHatEq}
    \hat{R}(u)\left(e^{u}E_0\otimes k_0+k_0^{-1}\otimes E_0\right)=
    \left(e^{u}k_0^{-1}\otimes E_0+E_0\otimes k_0\right)\hat{R}(u).
\ee
The operators $E_0\otimes k_0$ and $k_0^{-1}\otimes E_0$
are not separately coproducts of any element of the quantum group and, therefore, can mix different irreducible representations. Let us denote their matrix elements by $A$ and $B$ respectively, as follows
\be
    E_0\otimes k_0 |\lambda_m;\mu;i\rangle=\sum_{j=1}^{N}
    \sum_{\mu_j\in \Omega_{\lambda_j}}
    \sum_{i_j=1}^{\# \mu_j}
    A^{\lambda_j;\mu_j;i_j}_{\lambda_m;\mu;i}|\lambda_j;\mu_j;i_j\rangle,
\ee
\be
    k_0^{-1}\otimes E_0 |\lambda;\mu;i\rangle=
    -\sum_{j=1}^{N}
    \sum_{\mu_j\in \Omega_{\lambda_j}}
    \sum_{i_j=1}^{\# \mu_j}
    B^{\lambda_j;\mu_j;i_j}_{\lambda_m;\mu;i}|\lambda_j;\mu_j;i_j\rangle\,,
\ee
where we have assumed that on each of the irreducible submodules $V_{\lambda_j}$ we have picked a basis of eigenvectors of the Cartan subalgebra of the ``horizontal'' quantum group $U_q(\mathfrak{sl}(n))$, and denoted the $i$-th vector of weight $\mu$ in representation of highest weight $\lambda_j$ as $|\lambda_j;\mu;i\rangle$. The symbol $\#\mu_j$ above refers to the multiplicity of weight $\mu_j$ in the representation of highest weight $\lambda_j$.
Suppose the highest weights $\lambda_1$ and $\lambda_2$ both occur in the decomposition (\ref{tensorProd}) once.
Acting with both sides of (\ref{affineRootHatEq}) on a vector $|\lambda_1;\mu_1;i_1\rangle$ (i.e. the $i$-th vector of weight $\mu_1$ belonging to the irreducible representations of highest weight $\lambda_1$), and inspecting the coefficient in front of $|\lambda_2;\mu_2;i_2\rangle$ we find that
\be
\label{eigvalPermEq}
    f_{\lambda_1}(u)(A^{\lambda_2;\mu_2;i_2}_{\lambda_1;\mu_1;i_1} e^{u} -B^{\lambda_2;\mu_2;i_2}_{\lambda_1;\mu_1;i_1})=f_{\lambda_2}(u)(A^{\lambda_2;\mu_2;i_2}_{\lambda_1;\mu_1;i_1}-B^{\lambda_2;\mu_2;i_2}_{\lambda_1;\mu_1;i_1}e^{u}).
\ee
Since the ratio of eigenvalues does not depend on $\mu_1$, $\mu_2$, $i_1$, $i_2$, the quantity $\xi_{\lambda_1}^{\lambda_2}$ defined as
\be
    e^{i\xi_{\lambda_1}^{\lambda_2}}=\sqrt{\frac{A^{\lambda_2;\mu_2;i_2}_{\lambda_1;\mu_1;i_1}}{B^{\lambda_2;\mu_2;i_2}_{\lambda_1;\mu_1;i_1}}}
\ee
can also depend only on representations, but not on the particular choice of basis vectors within the representations (provided $B^{\lambda_2;\mu_2;i_2}_{\lambda_1;\mu_1;i_1}$ is nonzero).
The ratio of eigenvalues then reads
\be
    \frac{f_{\lambda_1}(u)}{f_{\lambda_2}(u)}=
    -\frac{\sinh(\frac{u}{2}-i\xi^{\lambda_2}_{\lambda_1})}{\sinh(\frac{u}{2}+i\xi^{\lambda_2}_{\lambda_1})}\,,
\ee
and the number $\xi_{\lambda_1}^{\lambda_2}$ can be obtained from the UV limit.
It may happen that for a given pair of representations, the coefficients $A$ and $B$ are identically zero for all pairs of vectors in these representations.
Then the ratio of two eigenvalues could be found as a product of ratios of eigenvalues of other representations occurring in (\ref{tensorProd}) with multiplicity one, which are connected to $\lambda_1$ and $\lambda_2$ by the action of the operators $E_0\otimes k_0$ and $k_0^{-1}\otimes E_0$. 

The analysis above is not applicable to the representations with multiplicities greater than one, as the equation (\ref{eigvalPermEq}) turns into an underdetermined system of linear equations for functions $M^{(\lambda_m)}_{ij}(u)$. We can only state that the functions $M^{(\lambda_m)}_{ij}(u)$ are rational functions of $e^u$.

 In the ``CFT'' approach, the eigenvalues of the $R$ matrix in the UV limit are related to the spectrum of dimensions of RCFT primaries,
which are labeled by weights of irreducible representations. These conformal dimensions read
\be
    \Delta_{\lambda}=\frac{(\lambda,\lambda+2\rho)}{2(k+g)}\,,
\ee
and the eigenvalues are expressed as
\be
\label{eigvalsCFT}
    f_{\lambda}=\epsilon_{\lambda}e^{i\pi \Delta_{\lambda}},
\ee
where $\epsilon_{\lambda}=\pm 1$.
For the $U_q(sl(3))$ at the generic level $k$ 
the decomposition (\ref{tensorProd}) reads
\be
\label{tensorProdAdAd}
    (1,1)\otimes(1,1)=
    (2,2)\oplus (3,0) \oplus (0,3)
    \oplus (1,1)_1 \oplus (1,1)_2 \oplus (0,0)\,,
\ee
and the corresponding RCFT conformal dimensions are given by
\be
    \Delta_{22}=\frac{8}{k+3}\,,
    \quad
    \Delta_{30}=\Delta_{03}=\frac{6}{k+3}\,,
    \quad
    \Delta_{00}=0\,,
    \quad
    \Delta_{11}=\frac{3}{k+3}\,.
\ee
The solution to the Yang-Baxter equation is fixed up to the choice of normalization function $s_1$. Normalized it so that
\be
    R_{1,1}^{1,1}(u)\equiv 1,
\ee
we obtain the following eigenvalues of $\hat{R}$ in the limit $u\rightarrow \pm \infty$:
\be
\label{eigvalOther}
    f_{22}=1\,,\quad
    f_{30}=f_{03}=-q^{\mp 2}\,,\quad
    f_{00}=q^{\mp 8}\,,
\ee
\be
\label{eigval11}
    f_{11}^{(\sigma)}=\sigma i q^{\mp 5}\,,\sigma\in \{-1,1\}\,.
\ee

In previous work \cite{Gepner:1992kx}, the functions $f_k(u)$ were computed for the R-matrices arising from RCFTs without multiplicities in the fusion rules. Their ratios read 
\be
\label{eigvalsGepner}
    \frac{f_{a}(u)}{f_b(u)}=\prod_{r=b}^{a-1}\frac{\sinh(i\xi_r-\frac{u}{2})}{\sinh(i\xi_r+\frac{u}{2})}\,,\quad
    \xi_r=\frac{\pi}{2}(\Delta_{r+1}-\Delta_{r}).
\ee
On the other hand, the inspection of our explicit solution yields the following expressions for the eigenvalues of the matrix $\hat{R}$:
\be
    f_{22}(u)=1\,,f_{30}(u)=f_{03}(u)=-\frac{\sinh(\frac{u}{2}-\frac{i\pi}{k+3})}{\sinh(\frac{u}{2}+\frac{i\pi}{k+3})}\,,
\ee
\be
    f_{00}(u)=\frac{\sinh(\frac{u}{2}-\frac{i\pi}{k+3})}{\sinh(\frac{u}{2}+\frac{i\pi}{k+3})}\frac{\sinh(\frac{u}{2}-\frac{3i\pi }{k+3})}{\sinh(\frac{u}{2}+\frac{3i\pi}{k+3})}.
\ee
Therefore, for the eigenvalues associated with the representations of multiplicity one there is an agreement between (\ref{eigvalsGepner}) and our results, despite the fact that the whole decomposition (\ref{tensorProdAdAd}) contains multiplicities, and there are coinciding eigenvalues. Thus it might be possible to generalize the CFT construction to the cases with multiplicities of both kinds.
 \section{Conclusions} \label{conclusions}
In this paper, we have investigated the unrestricted IRF model based on the affine Lie algebra $\mathfrak{su}(3)_k$, where the fluctuating variables residing on the lattice vertices belong to the set of integral weights $P$ (\ref{inw}) of the mentioned algebra. The admissibility conditions of the face configurations are defined by the adjoint representation of the corresponding finite Lie algebra $\mathfrak{su}(3)$. By utilizing the Vertex-IRF correspondence, we have determined the solutions for the trigonometric BWs of this model, which are given by the relation (\ref{vertexIRFeq}) and the quantum $R$ matrix elements (\ref{matrixREl5}).

In principle, the procedure applied here can be extended to solve any unrestricted IRF model based on a generic affine Lie algebra and a generic representation (defining the admissibility conditions) of this algebra by finding the corresponding quantum $R$ matrix.

As mentioned in the introduction, one of our aims in studying IRF models is to develop a method (or combinations of methods) through which one can find the BWs for more complex IRF models (e.g., those based on the extended current algebras discussed in \cite{Belavin:2023uqr, Baver:1996kf, Baver:1996mw}), and to explore further connections with CFTs. This research direction for IRF models is both intriguing and interesting, particularly considering the relevance of CFTs in current times. We plan to explore it further in future works.

\textbf{Acknowledgment:} J.R. thanks the organizers of the Workshop on Integrability 2023 held at the University of Amsterdam, where the result of this work was announced. The work of V.B. and J.R. is supported in part by \enquote{Program of Support of High Energy Physics}, Grant RA2300000222 by the Israeli Council for Higher Education.

\appendix

\section{Generators of the $U_q(\mathfrak{sl}(3))$ algebra} \label{generators}
 The generators of the $U_q(\mathfrak{sl}(3))$ quantum algebra (\ref{sl3qa}) in the adjoint representation can be written in the following manner. The Cartan generators are given by

\begin{equation} \label{apgenh1h2}
H_1=   \left(
\begin{array}{cccccccc}
 1 & 0 & 0 & 0 & 0 & 0 & 0 & 0 \\
 0 & -1 & 0 & 0 & 0 & 0 & 0 & 0 \\
 0 & 0 & 0 & 0 & 0 & 0 & 0 & 0 \\
 0 & 0 & 0 & 1 & 0 & 0 & 0 & 0 \\
 0 & 0 & 0 & 0 & -1 & 0 & 0 & 0 \\
 0 & 0 & 0 & 0 & 0 & 2 & 0 & 0 \\
 0 & 0 & 0 & 0 & 0 & 0 & 0 & 0 \\
 0 & 0 & 0 & 0 & 0 & 0 & 0 & -2 \\
\end{array}
\right), \quad
 H_2 = \left(
\begin{array}{cccccccc}
 1 & 0 & 0 & 0 & 0 & 0 & 0 & 0 \\
 0 & 2 & 0 & 0 & 0 & 0 & 0 & 0 \\
 0 & 0 & 0 & 0 & 0 & 0 & 0 & 0 \\
 0 & 0 & 0 & -2 & 0 & 0 & 0 & 0 \\
 0 & 0 & 0 & 0 & -1 & 0 & 0 & 0 \\
 0 & 0 & 0 & 0 & 0 & -1 & 0 & 0 \\
 0 & 0 & 0 & 0 & 0 & 0 & 0 & 0 \\
 0 & 0 & 0 & 0 & 0 & 0 & 0 & 1 \\
\end{array}
\right),
\end{equation}

the raising and lowering generators are
\begin{equation}
E_1=\left(
\begin{array}{cccccccc}
 0 & 1 & 0 & 0 & 0 & 0 & 0 & 0 \\
 0 & 0 & 0 & 0 & 0 & 0 & 0 & 0 \\
 0 & 0 & 0 & 0 & 0 & 0 & 0 & 0 \\
 0 & 0 & 0 & 0 & 1 & 0 & 0 & 0 \\
 0 & 0 & 0 & 0 & 0 & 0 & 0 & 0 \\
 0 & 0 & 1 & 0 & 0 & 0 & \frac{q^2+1}{q} & 0 \\
 0 & 0 & 0 & 0 & 0 & 0 & 0 & \frac{q^2+1}{q} \\
 0 & 0 & 0 & 0 & 0 & 0 & 0 & 0 \\
\end{array}
\right), \quad    
 E_2=  \left(
\begin{array}{cccccccc}
 0 & 0 & 0 & 0 & 0 & 1 & 0 & 0 \\
 0 & 0 & \frac{q^2+1}{q} & 0 & 0 & 0 & 1 & 0 \\
 0 & 0 & 0 & \frac{q^2+1}{q} & 0 & 0 & 0 & 0 \\
 0 & 0 & 0 & 0 & 0 & 0 & 0 & 0 \\
 0 & 0 & 0 & 0 & 0 & 0 & 0 & 0 \\
 0 & 0 & 0 & 0 & 0 & 0 & 0 & 0 \\
 0 & 0 & 0 & 0 & 0 & 0 & 0 & 0 \\
 0 & 0 & 0 & 0 & 1 & 0 & 0 & 0 \\
\end{array}
\right) ,
\end{equation}

\begin{equation}
F_1= \left(
\begin{array}{cccccccc}
 0 & 0 & 0 & 0 & 0 & 0 & 0 & 0 \\
 1 & 0 & 0 & 0 & 0 & 0 & 0 & 0 \\
 0 & 0 & 0 & 0 & 0 & 0 & 0 & 0 \\
 0 & 0 & 0 & 0 & 0 & 0 & 0 & 0 \\
 0 & 0 & 0 & 1 & 0 & 0 & 0 & 0 \\
 0 & 0 & 0 & 0 & 0 & 0 & 0 & 0 \\
 0 & 0 & 0 & 0 & 0 & 1 & 0 & 0 \\
 0 & 0 & \frac{q}{q^2+1} & 0 & 0 & 0 & 1 & 0 \\
\end{array}
\right), \quad F_2=  \left(
\begin{array}{cccccccc}
 0 & 0 & 0 & 0 & 0 & 0 & 0 & 0 \\
 0 & 0 & 0 & 0 & 0 & 0 & 0 & 0 \\
 0 & 1 & 0 & 0 & 0 & 0 & 0 & 0 \\
 0 & 0 & 1 & 0 & 0 & 0 & \frac{q}{q^2+1} & 0 \\
 0 & 0 & 0 & 0 & 0 & 0 & 0 & 1 \\
 1 & 0 & 0 & 0 & 0 & 0 & 0 & 0 \\
 0 & 0 & 0 & 0 & 0 & 0 & 0 & 0 \\
 0 & 0 & 0 & 0 & 0 & 0 & 0 & 0 \\
\end{array}
\right).
\end{equation}

\bibliographystyle{JHEP} 
\bibliography{refs} 
\end{document}